\documentstyle[twocolumn,floats,cite,citesupernumber,aps,epsfig]{revtex}

\begin{document}

\title{Precision Measurements of Stretching and Compression in Fluid Mixing}
\author{G. Voth$^{1}$, G. Haller$^{2}$, and J.P. Gollub$^{1,3,}$
\thanks{E-mail:jgollub@haverford.edu}}
\address{$^{1}$Department of Physics, Haverford College, Haverford\\
PA 19041, U.S.A.\\
$^{2}$Division of Applied Mathematics, Brown\\
University, Providence, RI 02912 U.S.A.\\
$^{3}$Department of Physics,\\
University of Pennsylvania, Philadelphia PA 19104, U.S.A}
\date{\today}

\maketitle

{\bf The mixing of an impurity into a flowing fluid is an
important process in many areas of science, including geophysical
processes, chemical reactors, and microfluidic devices.  In some
cases, for example periodic flows, the concepts of nonlinear
dynamics provide a deep theoretical basis for understanding
mixing\cite{Aref84,Ottino89,RomKedar90,RomKedar94,Beigie94,Giona99}.
Unfortunately, the building blocks of this theory, {\it i.e.} the
fixed points and invariant manifolds of the associated
Poincar\'{e} map, have remained inaccessible to direct
experimental study, thus limiting the insight that could be
obtained.  Using precision measurements of tracer particle
trajectories in a two-dimensional fluid flow producing chaotic
mixing, we directly measure the time-dependent stretching and
compression fields. These quantities, previously available only
numerically, attain local maxima along lines coinciding with the
stable and unstable manifolds, thus revealing the dynamical
structures that control mixing.  Contours or level sets of a
passive impurity field are found to be aligned parallel to the
lines of large compression (unstable manifolds) at each instant.
This connection appears to persist as the onset of turbulence is
approached.}

The relationship between the velocity field of a fluid flow and
the pattern formed by an impurity that it disperses can be
intricate. Even simple time-periodic flows in two dimensions can
produce chaotic mixing and complex distributions of material, in
which nearby fluid elements diverge strongly from each
other~\cite{Aref84}.  The fundamental processes involve a
combination of repeated stretching and folding of fluid elements
in combination with diffusion at small scales.   However, to
understand how the complex distributions of material actually
arise, it is important to determine the nonlinear maps that
connect the positions of fluid elements at different times, and to
show how these maps separate nearby elements. This requires more
precise and rapid measurement of flow fields than has been
accomplished previously.

Our work depends on high resolution measurements of particle
displacements in a two dimensional flow described later.
Approximately 800 fluorescent latex particles ($120~\mu {m}$ in
diameter) are suspended in the flow and followed by recording up
to 15,000 512x512 pixel images at 10 Hz in a typical run, or
40-180 images per period. The centroid of each of the 12,000,000
particles in the sequence of images is found to a precision of
about $40~\mu {m}$ (0.2 pixels). Particles found in sequential
images are then combined into tracks. Since the flow studied here
is time-periodic, we use conditional sampling, grouping together
particle positions in all images at the same phase relative to the
forcing. This process yields 100,000 precise particle positions at
each phase, velocities accurate to a few percent, spatial
resolution of 0.003 of the field of view, and time resolution of
about 0.01 periods.

We rely on recent theoretical work in which stable and unstable
manifolds are determined from finite-time velocity or displacement
data\cite{Miller97,Haller00,Haller01,Coulliette01}. Specifically,
from the measured velocity field, we determine the flow map
$\vec{x}^{\,\prime }=\vec{\Phi}(\vec{x},t_{0},\Delta t),$ a
function that specifies the destination vector $\vec{x}^{\,\prime
}$ at time $t_{0}+\Delta t$ of any fluid particle starting from
$\vec{x}$ at time $t_{0}$. (For $\Delta t$ equal to one period,
$\vec{\Phi}$ becomes the Poincar\'{e} map of the flow.)  The
stretching and compression fields are then obtained from its
gradients with excellent resolution. We identify the maximal
stretching experienced by a fluid element as the square root of
the largest eigenvalue of the right Cauchy-Green strain tensor,
$C_{ij},$ at that location: $C_{ij}=(\partial \Phi _{k}/\partial
x_{i})(\partial \Phi _{k}/\partial x_{j}),$ where summation is
implied over the repeated index $k=1,2$. Similarly, the maximal
compression at any location $\vec{x}$ is obtained as the
stretching for the backward-time flow map
$\vec{\Phi}(\vec{x},t_{0},-\Delta t).$  We note that the largest
finite-time Lyapunov exponent is given by the logarithm of the
stretching after division by $2\Delta t$. The stable and unstable
manifolds should coincide with local maxima of the stretching and
compression fields\cite{Haller01}.

To extract the flow map and its gradients, we first measure
particle velocities from trajectories using polynomial fitting.
The velocities are then interpolated onto a grid to obtain the
velocity as a function of space and phase. Numerical integration
of hypothetical particle trajectories from these velocity fields
produces flow maps with extremely high resolution. Although errors
in measured particle positions grow exponentially in chaotic
flows, the errors in the extracted invariant manifolds remain
small, as we confirm experimentally.

The two-dimensional flow is produced by density stratification and
time-periodic magnetic forcing\cite{Rothstein99}. A sinusoidal
electric current through a thin conducting fluid layer placed
above an array of permanent magnets generates a flow by means of
Lorenz forces. The fluid of interest is a 1 mm thick
non-conducting upper layer floating on the lower driven layer. The
fluids are glycerol-water mixtures, with the lower layer also
containing salt.  Though miscible, the two layers remain distinct
over the course of an experiment, and the flow stays essentially
two-dimensional.  The flow is a time-dependent vortex array,
spatially disordered in the work described here.  The flow is
15x15 cm, and all the figures in this paper show a central 10x10
cm region.  Typical forcing frequencies are 10-200 mHz, and
typical velocities are 0.05-1~cm/s. In some experiments, part of
the upper layer contains fluorescein dye, whose emission in the
visible under UV illumination is accurately proportional to the
local concentration.

The general behavior of this system has been described
elsewhere\cite{Rothstein99}.  After an initial transient, the
concentration field reaches a nearly steady state in which
stretching and folding balance diffusion in such a way that the
pattern recurs once per cycle of the forcing, except for a slow
overall exponential decay of contrast. This striking process may
be viewed in the supplementary animation available
on-line\cite{animations}.

There are two important control parameters. The Reynolds number
$Re=UL/\nu$ (based on the mean magnet spacing $L=2$~cm, rms
velocity $U$, and kinematic viscosity $\nu$) is typically between
10 and 200. The second parameter is the mean path length $p=UT/L$
in one forcing period $T$, which is typically in the range 0.5 to
10. Chaotic mixing is weak at the lower end of the ranges of $Re$
and $p$, where the unmixed elliptic regions are large, and mixing
grows stronger as $Re$ and $p$ are increased. Both parameters are
controlled by the forcing current, its frequency, and the fluid
viscosity. The flow becomes non-periodic or weakly turbulent in
the range $Re=100-150$, depending on $p$.

Figure 1 shows examples of one component of the velocity field for
a run at $Re=45$ and $p=1$. Both components are available as a
function of time, and an animation may be viewed
on-line\cite{animations}. The two fields shown in Figure 1 are
taken at equal time increments before and after the minimum of the
magnitude of the velocity field. Because these are not identical
(due to the non-zero $Re$), particles do not generally retrace
their paths and chaotic mixing occurs, despite the time reversal
symmetry of the forcing.

Studying mixing requires following fluid elements over extended
times. Fig. 2 shows the particle displacements over one period,
{\em i.e.} a Poincar\'{e} map. In these maps, lines are drawn from
the measured initial to final particle positions, and pseudocolor
is used to show the large and small displacements. Points at which
there is no net motion over a full period are called fixed points.
Both elliptic and hyperbolic fixed points may be seen. Figure 2(B)
shows a 4X enlargement of part of the Poincar\'{e} map. The fixed
points in this map have been marked. The hyperbolic fixed points
are saddles, with one axis of approach (stable manifolds) and one
axis of departure (unstable manifolds). In the remainder of this
paper we show how these fixed points and their stable and unstable
manifolds are related to the stretching and compression fields and
to the concentration field of an impurity such as a dye.

The key to this analysis is to compute the compression and
stretching fields. We begin (Fig. 3A) with an example of the
compression field, which is a measure of local convergence. It
consists of large values along relatively sharp lines, and much
smaller values between them. Its determination requires the
selection of a time interval $\Delta t$ for the map; here it is 3
periods. Much smaller values result in broader structures, because
less time is then available for compression to occur. Much larger
values of $\Delta t$ cause the flow map to develop structures that
are smaller than the spatial resolution. The optimum choice is a
balance of competing factors, but there is a fairly wide
acceptable range between 1 and 5 periods. An animation showing how
the compression field depends on $\Delta t$ is available
on-line\cite{animations}.

The dye concentration field is shown at the same phase (or time)
in Fig. 3B.  In Fig. 3C, we superimpose the compression field on
the dye image.  The dye visualization and particle tracking data
are measured in separate experiments at the same parameters.  We
observe that the level sets (contour lines) of the concentration
field line up with the lines of strong compression. Furthermore,
we find this to be the case at each instant or phase (see the
on-line animation\cite{animations}). Several authors have
described similar asymptotic alignment properties for numerically
computed two-dimensional time-periodic
flows\cite{RomKedar90,Giona99,Muzzio91,Pentek95}.

Now we discuss the correspondence between the lines of stretching
and compression, and the fixed points of the Poincar\'{e} map, as
shown in Fig. 4. Here we show the stretching field (blue) in
addition to the compression field (red). One can see that many of
the points where the two sets of lines cross correspond to
hyperbolic fixed points of the flow field. The lines of maximum
stretching and compression label the stable and unstable manifolds
of these fixed points\cite{Haller01}. Thus, we have the following
correspondence between the various objects: large stretching
$\leftrightarrow $ stable manifolds; large compression
$\leftrightarrow $ unstable manifolds. This demonstrated ability
to measure the locations of both the stable and unstable manifolds
as a function of time in complicated experimental flows opens new
possibilities to apply the insights of lobe
dynamics\cite{RomKedar90,RomKedar94} to practical mixing flows. An
animation\cite{animations} shows the superimposed stretching and
compression fields, which form homoclinic and heteroclinic tangles
of invariant manifolds.

We have also carried out this analysis at higher values of $Re$
and $p$, as shown in Fig. 5. Under these conditions, the fixed
points themselves are much harder to determine, because the
stretching and compression are then much greater. Our measurements
do not reveal any regular islands under these conditions. However,
the concentration field is still organized by the invariant
manifolds, with contours of constant concentration aligning with
the lines of large compression.  This is particularly dramatic in
a time-dependent animation\cite{animations}.

It is of interest to consider how the dynamics of mixing is
affected when the flow becomes weakly turbulent, i.e. nonperiodic.
Theoretical and numerical work indicates that the stretching and
compression fields continue to form well defined lines that
organize mixing\cite{Haller00}. To explore this issue
experimentally, we analyzed the stretching in a flow at $Re=115$,
$p=5$. Though $Re$ is only slightly elevated compared to the flow
in Fig. 5, the velocity field has bifurcated to a period doubled
state that repeats every second forcing period. We find that the
lines of the stretching and compression fields remain nearly
unchanged and continue to align with the dye pattern at each
instant. This suggests that the same will be true even in
non-periodic flows, and that there will therefore be a close
correspondence between the kinematics of chaotic and turbulent
mixing.

We have shown that lines of maximal stretching and compression
corresponding to invariant manifolds of flow maps can be
determined from precise experimental measurements of particle
trajectories. These special material lines, which emerge from
hyperbolic fixed points of the flow map, organize the evolution of
inhomogeneous impurities in the flow. We find that the dye contour
lines and the lines of maximal compression are locally parallel.
This is true at each instant in the time-dependent flow, and
continues to be the case at higher $Re$ where the fixed points
themselves are hard to determine.

It is striking to realize that the intricate dynamical processes
revealed here underlie mixing in many natural and artificial
flows.  It would be particularly interesting to predict the rate
of homogenization of an impurity from the stretching and
compression fields.  Theoretical models show how this may be
accomplished by extending this work to measure stretching over
longer times than is currently feasible\cite{Antonsen96}.
Eventually it may prove possible to utilize measurements such as
these to control mixing by modifying the behavior in regions that
are particularly active such as the region where the lines are
dense in Fig. 4.

\section{Acknowledgements}  We appreciate helpful
discussions with B. Eckhardt and I. Mezic.



\newpage

\begin{figure}[tb]
\begin{center}
\epsfig{file=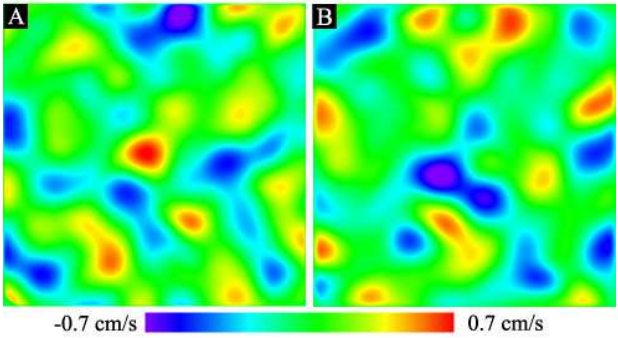,width=3.2in}
\end{center}
\caption{Maps of one component of the velocity field in the 2D
flow measured at two instants equidistant from the instant of
minimum flow ($p=1$, $Re=45$).   The fact that one image is not
the negative of the other is an example of the breaking of time
reversal symmetry required for chaotic mixing.}
\end{figure}

\begin{figure}[tb]
\begin{center}
\epsfig{file=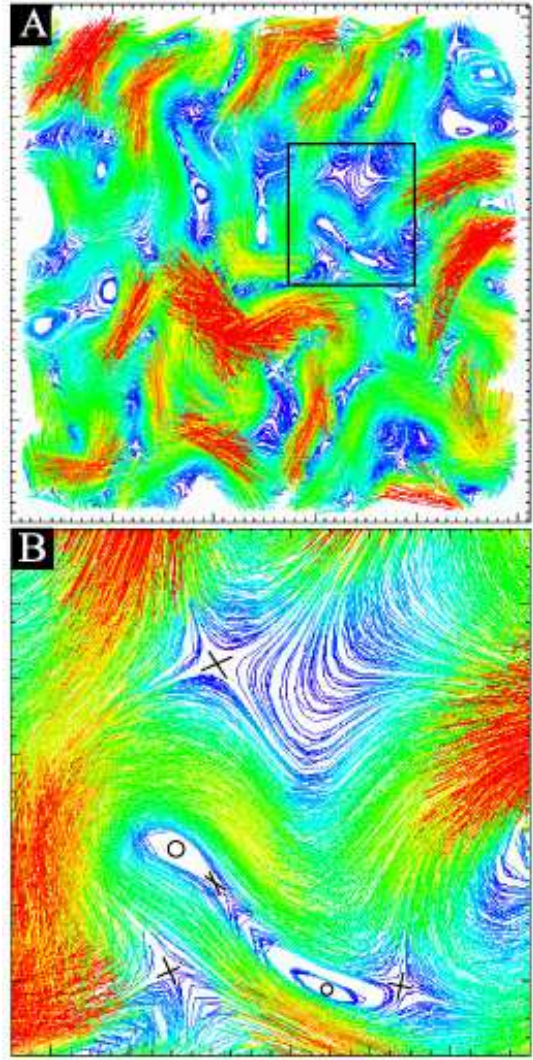,width=2.8in}
\end{center}
\caption{Poincar\'{e} map of the flow  at one phase of the forcing
($p=1$, $Re=45$). Lines connect the experimentally measured
initial coordinates of each particle with its coordinates one
period later, with blue and red designating small and large
displacements respectively. (A) Complete field of view. (B) A
close-up of the region in the box in (A), showing two elliptic
fixed points (circles), four hyperbolic fixed points (crosses),
and the trajectories near them. The hyperbolic point between the
two eliptic points is not fully resolved, but is clearly revealed
by the manifolds in Fig. 4.}
\end{figure}

\begin{figure}[tb]
\begin{center}
\epsfig{file=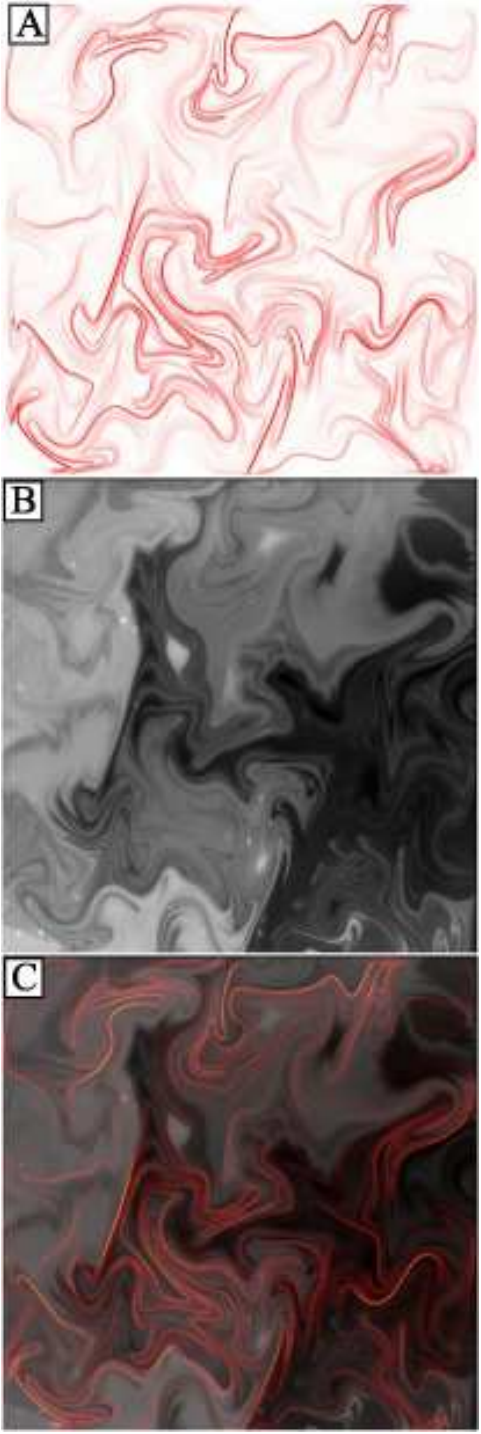, width=2.5in}
\end{center}
\caption{(A) Compression field calculated from trajectories
obtained by integration of the measured velocity field. (B)
Corresponding dye image showing the concentration after about 30
periods but at the same phase as in (A). (C) Superposition of the
compression field with the corresponding dye image; the contour
lines of the concentration field are aligned with the lines of
large compression. This is from the same flow at the same phase as
Fig. 2.}
\end{figure}

\begin{figure}[tb]
\begin{center}
\epsfig{file=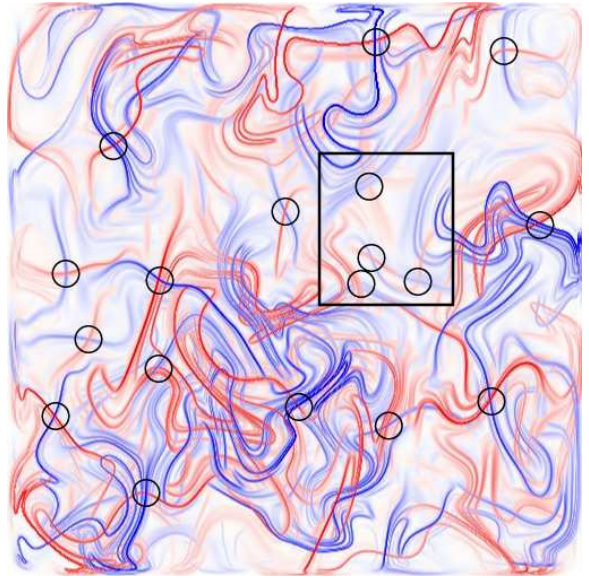,width=3.0in}
\end{center}
\caption{Lines of the stretching field (blue) and compression
field (red), with some of the hyperbolic fixed points of the flow
map marked by circles. Conditions match Figs. 2 and 3.  The lines
are respectively the stable and unstable manifolds of the fixed
points of the Poincar\'{e} map. The box marks the region which is
shown in Fig. 2(B).  Note that some of the fixed points have only
very weak stretching and compression lines associated with them.
An online animation{\protect\cite{animations}} shows the time
dependence of the manifolds, which allows much greater insight
into the dynamics of the mixing process.}
\end{figure}

\begin{figure}[tb]
\begin{center}
\epsfig{file=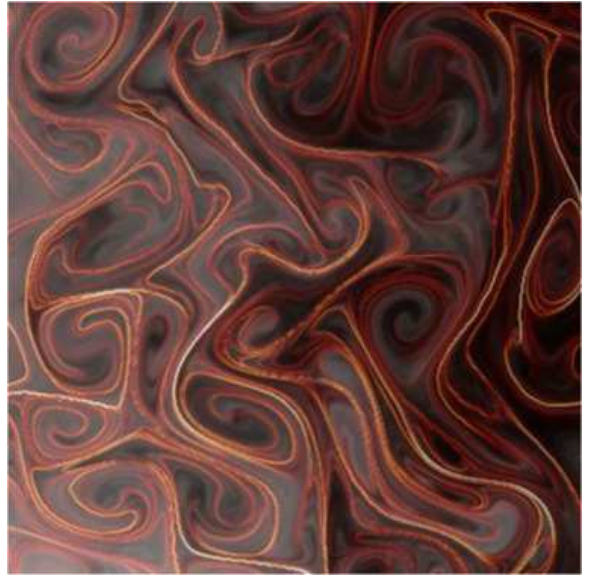,width=3.0in}
\end{center}
\caption{Superposition of the compression field with a dye image
at $Re=100$ and $p=5$.  The pattern here resembles an array of
random vortices, and the compression field lines are again found
to align with the concentration field at each instant.}
\end{figure}

\end{document}